# Tail-states induced semiconductor-to-conductor like transition under sub-bandgap light excitation in the zinc-tin-oxide photo-thinfilm transistors

Soumen Dhara[1,a], Kham M. Niang[2], Andrew J. Flewitt[2], Arokia Nathan[2], and Stephen A. Lynch[1]

We report on a giant persistent photoconductivity (PPC) induced semiconductor-to-conductor like transition in zinc-tin-oxide (ZTO) photo-thinfilm transistors (TFT). The active ZTO channel layer was prepared by remote-plasma reactive sputtering and possesses an amorphous structure. Under sub-bandgap excitation of ZTO with UV light, the photocurrent reaches as high as ~$10^{-4}$ A (a photo-to-dark current ratio of ~$10^7$) and remains close to this high value after switching off the light. During this time, the ZTO TFT exhibits gigantic PPC with long-lasting recovery time, which leads the ZTO compound to undergo a semiconductor-to-conductor like transition. In the present case, the conductivity changes over six orders of magnitude, from ~$10^{-7}$ to 0.92 $\Omega^{-1}$cm$^{-1}$. After UV exposure, the ZTO compound can potentially remain in the conducting state for up to a month. The underlying physics of the observed PPC effect is investigated by studying defects (deep-states and tail-states) by employing a discharge current analysis (DCA) technique. Findings from the DCA study reveal direct evidence for the involvement of sub-gap tail-states of the ZTO in the giant PPC, while deep-states contribute to mild PPC.

Coupling, or interaction between constituent materials, in ternary/quaternary compounds or hybrid composites has the potential to unlock exceptional properties which are superior to those of the pristine binary materials. Furthermore, the bandgap energy can be easily tuned by changing the compositional stoichiometry of the ternary/quaternary oxide semiconductors according to requirements. Recently, amorphous compound oxide semiconductors (AOS) have unlocked new electronic and optoelectronic functionality with the development of transparent and flexible devices[1-6]. The initial breakthrough came through the first demonstration of indium-gallium-zinc oxide (IGZO) thinfilm transistors (TFTs) by the Hosono group in 2004[7]. IGZO based TFTs have been used in flat panel displays[8] and flash memory applications[9] due to its high electron mobility, strong light sensitivity, and ultralow power consumption. However, many research laboratories have recently ramped up research on alternative AOS TFT materials to avoid over-reliance on indium[2,10,11]. Similar to IGZO, a simple ternary oxide, zinc-tin-oxide (ZTO), is also very sensitive to light and highly conductive[12]. In addition, ZTO has better chemical stability against oxidation and chemical etching[13]. Recently, TFTs made with ZTO active channels have been demonstrated, and show promising electrical characteristics[6,14-16]. The reported results/performance of ZTO based TFTs indicate that it could be a good competitor to IGZO based TFTs. However, the photo-electrical properties of ZTO TFTs have been less well-explored, and there are very few reports in the literature[17-19]. Oxide semiconductors suffer from one critical issue, namely persistent photoconductivity (PPC), which hinders successful commercial applications in photosensing. In order to resolve this issue, various materials and approaches have been introduced and studied over the past few years[18,20-22]. However, a straightforward and effective technology applicable to all AOS has yet to develop. Although, it is believed that oxygen vacancies are responsible for the PPC effect, direct evidence for this has yet to emerge.

[1]School of Physics and Astronomy, Cardiff University, Cardiff CF24 3AA, United Kingdom. [2]Electrical Engineering Division, Department of Engineering, University of Cambridge, Cambridge CB3 0FA, United Kingdom. [a]Present address- Department of Physics and Mathematics, Sri Sri University, Cuttack- 754006, India, Correspondence and requests for materials should be addressed to, SD (email: soumen5484@yahoo.co.in) or SAL (email: LynchSA@cardiff.ac.uk)

Consequently, the lack of in-depth understanding of PPC in binary and ternary oxides has impeded its useful exploitation in optoelectronics. Therefore, it is very important to conduct a detailed study on the photoexcited transport and sensing properties of the ZTO for the development of next-generation flexible optoelectronic devices.

In this work, we investigated photoexcited electrical transport characteristics of ZTO TFTs with different concentrations of tin for a range of different photon energies. We observed giant PPC with a very high value of the photo-to-dark current ratio (sensitivity) and a semiconductor-to-conductor like transition. Under the illumination of UV light at 365 nm, with photon energy just below the band-to-band excitation energy of ZTO, the sensitivity reaches as high as ~$10^7$. With the exposure of UV light for 15 min, we observe a pronounced semiconductor-to-conductor like transition in the ZTO material, with high photosensitivity and giant PPC (lasting up to a month). To investigate further the observed giant PPC, we studied the material's photo-response, by illuminating TFTs at different photon energies in the sub-gap region. We extracted the ZTO defect density, and density of sub-gap states (DOS), by employing a discharge current analysis (DCA) technique. The DCA results show that ZTO TFTs with higher tin concentration exhibit higher sub-gap tail states. A correlation between photoexcitation energy and DOS distribution is discussed in detail, leading to a direct evidence of the involvement of tail-states of the ZTO in the observed giant PPC.

## Results and Discussion

The active semiconductor channel of both sets of fabricated TFTs (33ZTO, contains 33 at.% of tin and 50ZTO, contains 50 at.% of tin) possess amorphous crystal structure, as expected. The surface morphology of the ZTO thinfilms is very smooth and presents a homogenous surface with a root mean square (RMS) surface roughness of 1.5 nm, measured from AFM images (data not shown). Details of microstructural and morphological characterization results can be found in the previous report[23]. The electrical measurement data of ZTO TFT shows the typical output characteristics ($I_{DS}$ vs. $V_{DS}$) and transfer characteristics ($I_{DS}$ vs. $V_{GS}$) of a field effect transistor, with distinct linear and saturation regions (Fig. 1). Several important characteristic parameters of the ZTO TFTs are extracted from the gate transfer graph for both the 33ZTO and 50ZTO TFTs. Based on standard field effect transistor theory[24], the field effect mobility ($\mu_{FE}$) and sub-threshold slope ($SS$) are extracted from a linear fit to the experimental $I_{DS}$ vs. $V_{GS}$ data at $V_{DS} = 0.1V$ in the linear region and log-scale plot, respectively using following equations,

$$\mu_{FE} = \frac{\frac{\partial I_{DS}}{\partial V_{GS}}}{C_{ox}(W/L)V_{DS}} \quad (1)$$

$$SS = \frac{dV_{GS}}{d\log(I_{DS})} \quad (2)$$

where $C_{ox}$, $W$, $L$, $I_{DS}$, $V_{DS}$ and $V_{GS}$ are gate oxide capacitance (~18 nF/cm$^2$), channel width (1900 μm), channel length (50 μm), drain-to-source current, drain-to-source voltage and gate-to-source voltage, respectively. The estimated $\mu_{FE}$ and $SS$ are found to be 9.07 cm$^2$/V.s and 0.92 V/dec, with a threshold voltage ($V_{th}$) of -0.1 V for the 33ZTO TFT. In the 50ZTO TFT, which contains a higher tin concentration, it possesses lower $\mu_{FE}$ of 6.07 cm$^2$/V.s and $SS$ of 0.65 V/dec with a $V_{th}$ of -3.9V. Here we note that the magnitude of the $V_{th}$ is estimated from the gate transfer data measured at a voltage sweep rate of 0.33 V/s. In oxide-based semiconductor TFTs, measurement of $V_{th}$ may be affected by voltage sweep rate due to the charge trapping effect[25,26]. The lower value of mobility in 50ZTO indicates that the density of charge trapping centers is higher in 50ZTO than the 33ZTO TFT. The estimated magnitudes of each of these parameters are comparable or slightly better than the rival IGZO based TFTs and ZnO or SnO$_2$ based TFTs [7,27-29]. So, both the TFTs can operate in the ON state at zero gate bias. However, the device requires a negative gate voltage to switch off the TFTs which increases the power consumption. Ideally, to reduce power consumption, it would be necessary to achieve a positive threshold voltage close to zero voltage.

To study the photoconductivity of semiconducting material, the first requirement is to know the exact bandgap energy of that material. The optical bandgap energy ($E_g$) of the ZTO thinfilm is estimated from its UV-visible transmission spectrum, as shown in Fig 2(a). Transmittance data of the ZTO thinfilms shows a highly transparent film with transparency above 85% in the visible-to-NIR region. The bandgap energy is calculated from the standard Tauc plot relation assuming a direct allowed transition. A linear fit to the ($\alpha h\nu$)$^2$ vs. energy ($h\nu$) and calculated $E_g$ are shown in Fig. 2(b). The estimated bandgap energies are found to be 3.52 eV and 3.66 eV for the ZTO thinfilms with 33 at.% and 50 at.% of tin concentrations, respectively. This provides evidence that ZTO thinfilms with a higher concentration of tin show larger bandgap than the case of lower tin concentrations, according to the Burstein-Moss effect. The electrical characterization data of the TFTs and Hall effect measurement of the ZTO thinfilms (data not shown, see Ref.# 23) show higher carrier concentration for higher Sn content sample thus supporting the bandgap widening hypothesis according to the Burstein-Moss effect. The estimated order of magnitude of carrier concentration of the ZTO thinfilms from Hall Measurement is $10^{17}/cm^3$. Furthermore, $E_g$ appears in a range between the bandgap energies of crystalline ZnO (~3.37 eV)[30] and SnO$_2$ (~4.05 eV)[31] films, as expected.

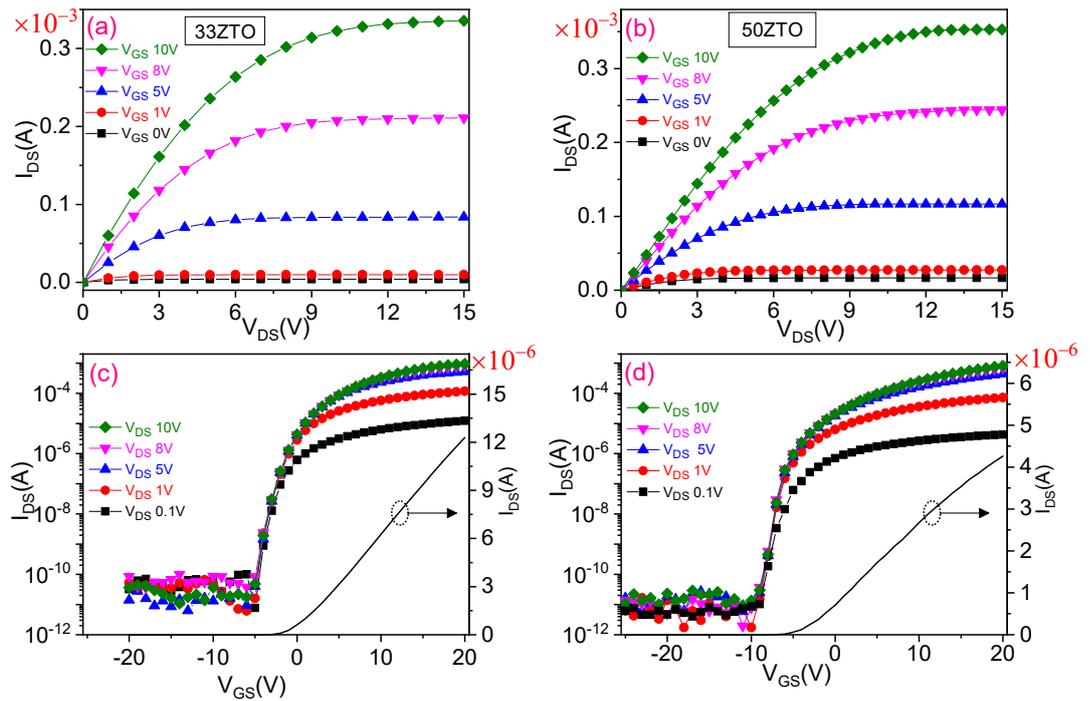

**Figure 1.** The output characteristics ($I_{DS}$ vs. $V_{DS}$) of ZTO TFTs for the sample (a) 33ZTO and (b) 50ZTO measured at different $V_{GS}$. In (c) and (d), the transfer characteristics ($I_{DS}$ vs. $V_{GS}$) of ZTO thinfilm transistors for the sample 33ZTO and 50ZTO, respectively. In both the TFTs, active semiconductor channel W/L ratio was fixed to 38.

…………………………………………………………………………………………………………………………...

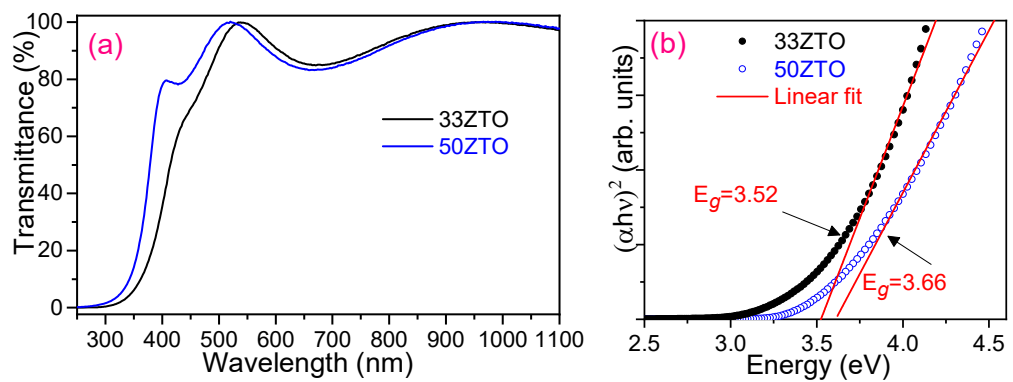

**Figure 2.** (a) UV-Vis transmission spectra and (b) calculation of the bandgap energy of the ZTO thinfilms on Corning glass substrate with an elemental composition identical to semiconductor channels of 33ZTO and 50ZTO TFTs.

…………………………………………………………………………………………………………………………...

To study the photoexcited transfer characteristic of ZTO TFTs, the device was exposed to UV light at 365 nm (intensity ~40 mW/cm$^2$) for 5 min before measuring $I_{DS}$ vs. $V_{GS}$ data. In order to concentrate on the defect-induced photoconductivity study of the ZTO TFTs, the photon energy was limited to below the bandgap energy of ZTO (3.52 - 3.66 eV). Under illumination, the photogenerated drain-to-source current of ZTO TFTs increases from a few picoamps to miliamps in the accumulation region, as shown in Fig. 3(a and b) and operates as a photo-TFT. In case of the 33ZTO TFT, at a gate bias of -6 V, the drain-to-source current increases from $6.2 \times 10^{-12}$ A to $4.7 \times 10^{-5}$ A, leading to a sensitivity (defined as the photo-to-dark current ratio at the same drain and gate voltages) up to $10^6$. Similarly, in the 50ZTO TFT at a gate bias of -10 V, the drain-to-source current increases from $1.7 \times 10^{-12}$ A to $1.3 \times 10^{-4}$ A, leading to a sensitivity of $10^7$. The sensitivity vs. gate voltage plot shows a strong

Submitted to 'Nature Scientific Reports'

gate-tuning effect, where the sensitivity gradually increases with decreasing applied gate bias, and saturates at a gate bias below the flatband voltage ($V_{FB}$). Furthermore, upon light illumination, a large negative shift in the threshold voltage ($\Delta V_{th} > 16$ V) is observed from both the TFTs and the devices remain in the ON state up to -20 V (measurement limit) of gate bias. The negative shift in the threshold voltage was observed previously in most of the AOS, where the magnitude of the negative shift depends on the light intensity and light stress time[20,32-34]. Both the TFTs exhibit excellent photoconductivity, with a very high value of sensitivity, which is one of the important criteria for development of the photo-TFT based smart UV sensors.

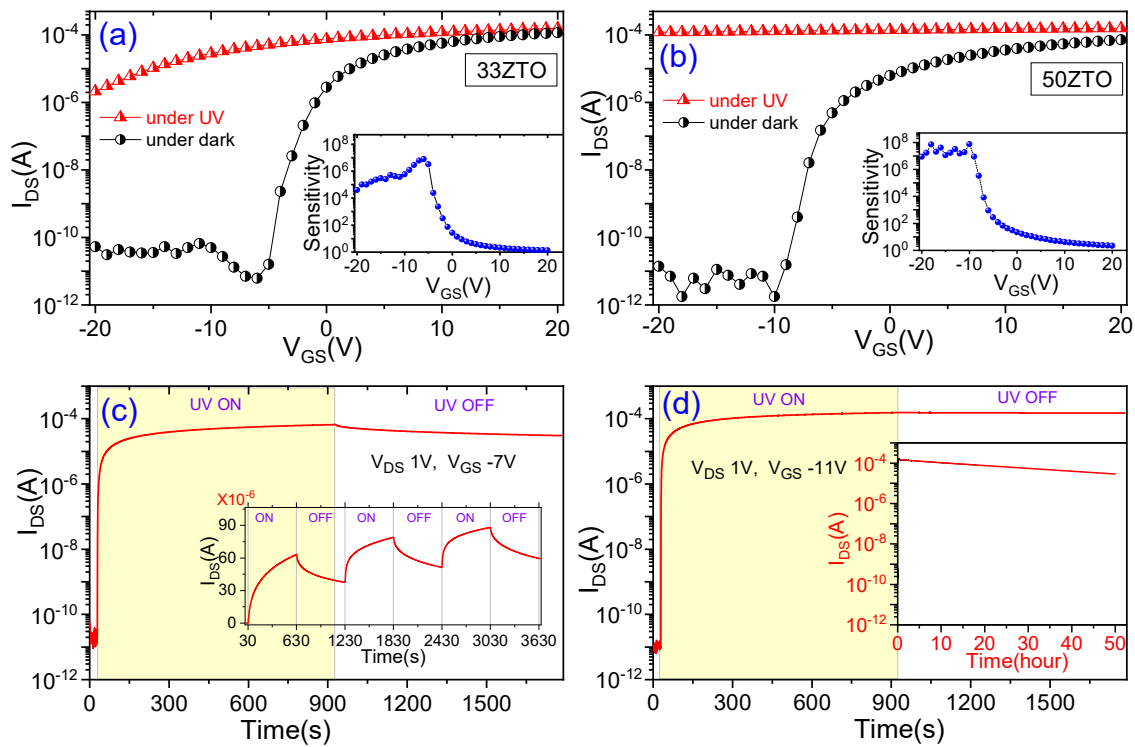

**Figure 3.** The transfer characteristics of (a) 33ZTO and (b) 50ZTO TFTs at $V_{DS} = 1$V measured in dark and under UV light (at 365 nm) exposure. The sample was exposed to UV light for 5 min at an intensity of 40 mW/cm² before starting the photocurrent-voltage measurement. Insets show the sensitivity (photo-to-dark current ratio) as a function of $V_{GS}$. In (c) and (d), the transient response of the drain-to-source current for ZTO photo-TFTs under a single UV light pulse of 15 min for the sample 33ZTO and 50ZTO, respectively. The inset in (c) shows photoresponse of 33ZTO under a periodic UV pulse (10 min). The inset in (d) shows extended measurement (up to 50 hrs) of drain-to-source photocurrent after switching off the UV light.

…………………………………………………………………………………………………………………...

Next, we studied the time response of the photo-TFTs by measuring transient drain-to-source current in the accumulation region in light ON and OFF conditions, as shown in Fig. 3(c and d). Upon UV illumination (at 365 nm), the photocurrent increases exponentially with time. However, in the present case for both the TFTs, the photocurrent neither saturates to a high current value nor does it reset to the initial dark current reading after 15 min of light exposure followed by no further illumination. After 15 min of decay measurements (after switching off the light), the photocurrent decreases to 54.4% of the maximum value for the 33ZTO TFT and only 2.9% for the 50ZTO TFT (which is an exceptionally slow recovery process). Therefore, both the TFTs exhibit a giant PPC effect with long-lasting recovery time. Transient photocurrent measurement under a periodic UV illumination pulse (10 min), as shown in the inset of Fig. 3(c) further confirms the repeatability of the observed long-lasting PPC in the ZTO TFTs. The maximum photocurrent in the next cycle is increased, compared to the previous cycle, because of the incomplete increase and recovery of the photocurrent during the measurement cycle. To estimate the reset time (defined as the time required for recovery to 37% (1/e) of the maximum photocurrent), decay measurements were performed for longer time periods. Extended measurement of drain-to-source photocurrent (up to 50 hr) for the 50ZTO TFT is shown as an inset in Fig. 3(d). The reset time is estimated to be 55.8 min and 29.7 hrs for the 33ZTO and 50ZTO TFTs, respectively. Here we note that 50ZTO TFT exhibits a gigantic PPC with recovery time ~30 times larger than the 33ZTO TFT. The mathematical extrapolation of the photocurrent decay data for the 50ZTO TFT indicates that this persistent photocurrent can be potentially maintained for up to a month. This long-lasting PPC far exceeds all previously reported results from other AOS[18,20,21]. We conclude from this, that tin concentration in the ZTO ternary oxide modulates the photosensitivity as well as PPC recovery time. The observed long-lasting PPC at a sub-gap photoexcitation implies that most of the excess electrons are generated from optically irreversible states. These irreversible states could be sub-gap deep defects states or tail states. In AOS, oxygen vacancy related defects ($V_o$) are very common. These defects become ionised under optical illumination ($V_o \rightarrow V_o^+/V_o^{2+} + e^-/2e^-$) providing free electrons[20,35,36]. It is believed that PPC is associated with these ionised oxygen defects, which oppose the recovery process by charge localisation inducing a slowdown in the reaction, $V_o^{2+} + 2e^- \rightarrow V_o$. However, in the present case, there is a vast difference in the long-lasting PPC between 33ZTO TFT and 50ZTO TFT, which indicates that in addition to oxygen vacancy defects, another irreversible state is also responsible. This result agrees with the XPS study (discussed later) which reveals that oxygen vacancy is higher in 33ZTO than the 50ZTO.

The strong PPC with a high value of sensitivity (~$10^7$) in the 50ZTO TFT causes the semiconducting channel to become an excellent conductor. In the present case, the conductivity of the 50ZTO compound in the dark is estimated as ~$10^{-7}$ $\Omega^{-1}$cm$^{-1}$. Upon UV exposure, conductivity increases to 0.92 $\Omega^{-1}$cm$^{-1}$, which is comparable to the conductivity of conductors like semi-metals, and remains at a high value for a long time after switching off the UV light After 50 hrs the conductivity was measured to be 0.16 $\Omega^{-1}$cm$^{-1}$. The UV exposure thus causes the 50ZTO sample to undergo a semiconductor-to-conductor like transition. We believe the highly photo-responsive characteristic of the ZTO semiconductor, combined with the semiconductor-to-conductor transition, may facilitate the favourable exploitation of PPC in the next generation smart optoelectronic devices.

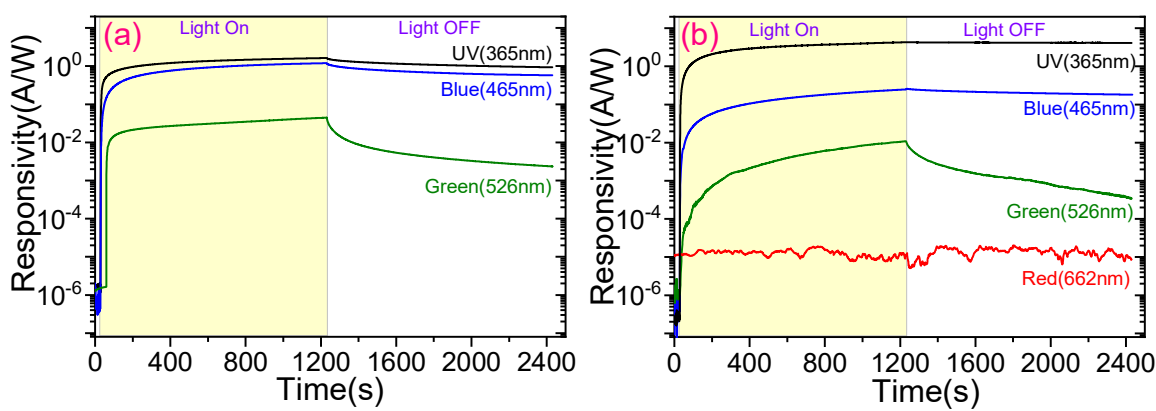

**Figure 4.** Transient response of responsivity (photocurrent generated per unit of applied light intensity) of the ZTO TFTs for the sample (a) 33ZTO and (b) 50ZTO measured at different photoexcitation energy: UV (at 365 nm); blue (at 465 nm); green (at 526 nm); red (at 662 nm).
……………………………………………………………………………………………………………...

To investigate further the observed strong PPC effect, we studied photoresponse of the TFTs in light ON and OFF conditions, for different photon excitation energies in the sub-gap region. For a comparative study, the measured photocurrent is converted to the responsivity by normalising against the light intensity of the specific light source. Figure 4 and Table 1 show the transient responsivity for the 33ZTO and 50ZTO TFTs measured under the illumination of UV (at 365 nm), blue (at 465 nm), green (at 526 nm) and red (at 662) lights. The transient photocurrent data under the illumination of blue and green light show similar characteristics to the responsivity measured under UV illumination. Under illumination by red light with photon energy far below the bandgap of 50ZTO, no notable photo-generation of charge carriers appears (Fig. 4b) due to an absence of sufficient activation of defect states. Similar behaviour for the 33ZTO under illumination by red light is expected. More interestingly, the PPC effect improves when the active channel is illuminated by blue/green light, which means there is a comparatively faster photocurrent decay during recovery. Under excitation by blue light, the reset times are 21.2 min and 70.1 min for the 33ZTO and 50ZTO TFTs respectively, which is far better than the case of photoexcitation with UV light. The reset time is further improved under excitation by green light (53 sec for 33ZTO and 90 sec for 50ZTO). The maximum responsivity for the 33ZTO after 20 min of light exposure gradually decreases with a decrease in photon energy, as expected (1.592 A/W under UV, 1.197 A/W under blue and 0.045 A/W under green). In contrast, the 50ZTO TFT responsivity is 4.350 A/W at UV and then drastically reduces with decreasing photon energy (0.259 A/W at blue and 0.011 A/W at green). It is well accepted that Vo defects are responsible for the slow decay process in the oxide semiconductors and decay times are comparable at different excitation energies. Our experimental results are also in agreement with this mechanism. Much faster comparative reset time can be expected under illumination with light of photon energy above the bandgap energy of the ZTO. Faster reset times of the order of few sec to several sec have been reported by different research groups from the high quality ZTO thinfilms at wavelength <350 nm (photon energy is higher than the bandgap energy of the ZTO)[19,37]. Relatively low sensitivity and slower photoresponse were observed when illuminated with light at wavelength 450 nm.

| Sample | Excitation Wavelength (nm) | Sensitivity | Responsivity (A/W) | Reset time (min) |
|---|---|---|---|---|
| **33ZTO** | 365 | $10^6$ | 1.592 | 55.8 |
|  | 465 | - | 1.197 | 21.2 |
|  | 526 | - | 0.045 | 0.9 |
| **50ZTO** | 365 | $10^7$ | 4.350 | 1782.2 |
|  | 465 | - | 0.259 | 70.1 |
|  | 526 | - | 0.011 | 1.5 |

**Table 1:** Extracted parameters from the photoresponse characteristics (Fig. 4) of the ZTO thinfilms.

………………………………………………………………………………………………………………...

It is observed that the photoresponse of the amorphous oxide semiconductor TFTs operated in negative bias under illumination stress (NBIS) depends very much on the photon energy and the intensity (mW/cm$^2$) of the light source. In a report by Chowdhury *et al.*[38], the well-performing and stable IGZO TFTs under NBIS with a 365 nm wavelength and only 0.7 mW/cm$^2$ of light intensity, exhibited a $V_{th}$ shift as high as ~ -10 V is reported. Fernandes *et al.*[39] reported a $V_{th}$ shift of -3.0 V under NBIS with a wavelength of 420 nm from a ZTO TFT annealed at 300 °C. It is likely that the light intensity is relatively small (the actual value is not mentioned) in this case but enough to cause a $V_{th}$ shift after 1 hr of exposure. In the present case, the photoresponse measurement under UV was carried out at a light intensity of 40 mW/cm$^2$, which is considered as a high value. In addition, ZTO thinfilms were grown at room temperature with low sputtering power. After a high temperature annealing at 500 °C the resulting trap density is of the order of $10^{12}$ cm$^{-2}$ eV$^{-1}$ which is still higher than the trap density present in the samples prepared by Fernendes *et al.*. This suggests that the 500 °C annealed devices do not necessarily result in better/more stable devices.

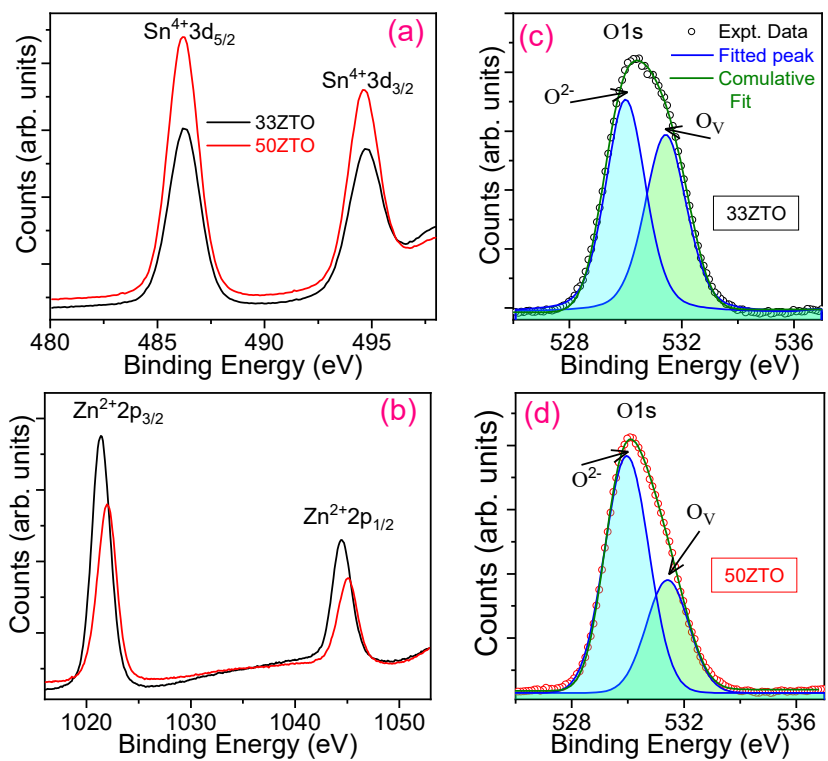

**Figure 5.** XPS binding energy spectra of the ZTO thinfilms with an elemental composition identical to semiconductor channel of 33ZTO and 50 ZTO TFTs for the core level of: (a) Sn3d, (b) Zn2p and (c and d) O1s. Individual component (lattice oxygen and oxygen vacancy) in the broad and asymmetric O1s spectrum is identified after fitting with multiple Gaussian peaks. Solid green line represents the combined fitting of the sum of all the individual components (blue line).

………………………………………………………………………………………………………………...

To analyse the relationship between the observed photocurrent features and oxygen vacancy content, we thoroughly investigated the XPS spectra for the O1s core level. Figure 5 shows the X-ray photoelectron spectroscopy (XPS) binding energy spectra for the core levels of Sn3d, Zn2p and O1s measured from 33ZTO and 50ZTO thinfilms. The XPS data for the Sn3d core level exhibits a peak doublet at 486.3 and 494.8 eV corresponding to $Sn^{4+}3d_{5/2}$ and $Sn^{4+}3d_{3/2}$, respectively[40]. Similarly, for the Zn2p core level, a doublet with peaks centered at 1021.4 and 1044.4 eV is observed, corresponding to $Zn^{2+}2p_{3/2}$ and $Zn^{2+}2p_{1/2}$, respectively. We noted that the integrated peak area of the Sn3d doublet increased from 33ZTO to 50ZTO thinfilms, whereas it is reduced for the Zn2p doublet, as expected due to increase in tin concentration. The broad and asymmetric peak of the O1s core level was deconvoluted, giving two individual components with peaks centered at 530.0 and 531.4 eV. The lower binding energy peak, $O^{2-}$, represents oxygen ions connected with metal ions, and the higher binding energy peak, $O_v$, is associated with oxygen ions that are in oxygen deficient regions within the matrix of ZTO ( oxygen vacancy)[41]. The trace of the absorbed water moisture ($H_2O$, appears at ~532.7 eV) was checked during deconvolution and no such peak observed. As the ZTO thinfilms underwent a post-growth annealing at a higher temperature (at 500 °C), there is also a remote possibility of the presence of water moisture on the surface of the thinfilms. When the spectra of the O1s core level for 33ZTO and 50ZTO are compared, the 33ZTO thinfilm shows higher intensity for oxygen vacancies. The relative oxygen vacancy content is estimated from the integrated peak area of $O^{2-}$ and $O_v$ peaks, and it changes from 46.0% to 32.5% with the increase of tin concentration from 33 at.% to 50 at.%. The XPS data reveals that 33ZTO contains more $V_o$ states than the 50ZTO thinfilm. The oxygen vacancies in the amorphous oxides are believed to be responsible for generating carriers in the depletion region and have a strong influence on the conductivity[28,42]. Here we note that, deconvolution of spectra of Zn2p and O1s core levels shows no traces of hydroxyl or hydrogen on the surface of the sample. Incorporating dopant atoms directly into the lattice of the semiconductor oxide by substitution is the preferred method of doping, and so we would expect XPS measurements of the ZTO system to show evidence of this. The ratio of atomic percentage of individual elements (Zn:Sn:O) is 0.268:0.139:0.593 for 33ZTO and 0.207:0.215:0.578 for 50ZTO. Assuming substitutional doping (Sn replaces Zn atom), the above measured ratios indicate ~1.5% surplus of zinc atoms present in 50ZTO sample. Therefore, we can say that there might be ~1.5% of zinc interstitial defects ($Zn_i$) present in the 50ZTO sample which cannot be ignored. The formation energy of neutral $Zn_i$ is 6.95 eV[43], which is the second lowest energy amongst all the other commonly found native defects in ZnO. This supports the hypothesis for a high probability of $Zn_i$ defects formation. In common with the $V_o$ defects, $Zn_i$ also acts as trap centers under voltage bias. The XPS result explains why we observe higher mobility in 33ZTO rather than the 50ZTO. We note however, that if oxygen vacancies are only responsible for the observed photoresponse features in ZTO TFTs (here, photon energy is not sufficient to create band-to-band electron-hole pairs), higher photocurrents and long-lasting PPC would be expected in 33ZTO rather than 50ZTO, which is the opposite of what we observe. Therefore, considering only the involvement of oxygen vacancy defects, which are usually distributed in the sub-gap region of 0.5-1.3 eV below the conduction band (CB), fails to explain the observed photoresponse and PPC in the ZTO TFTs.

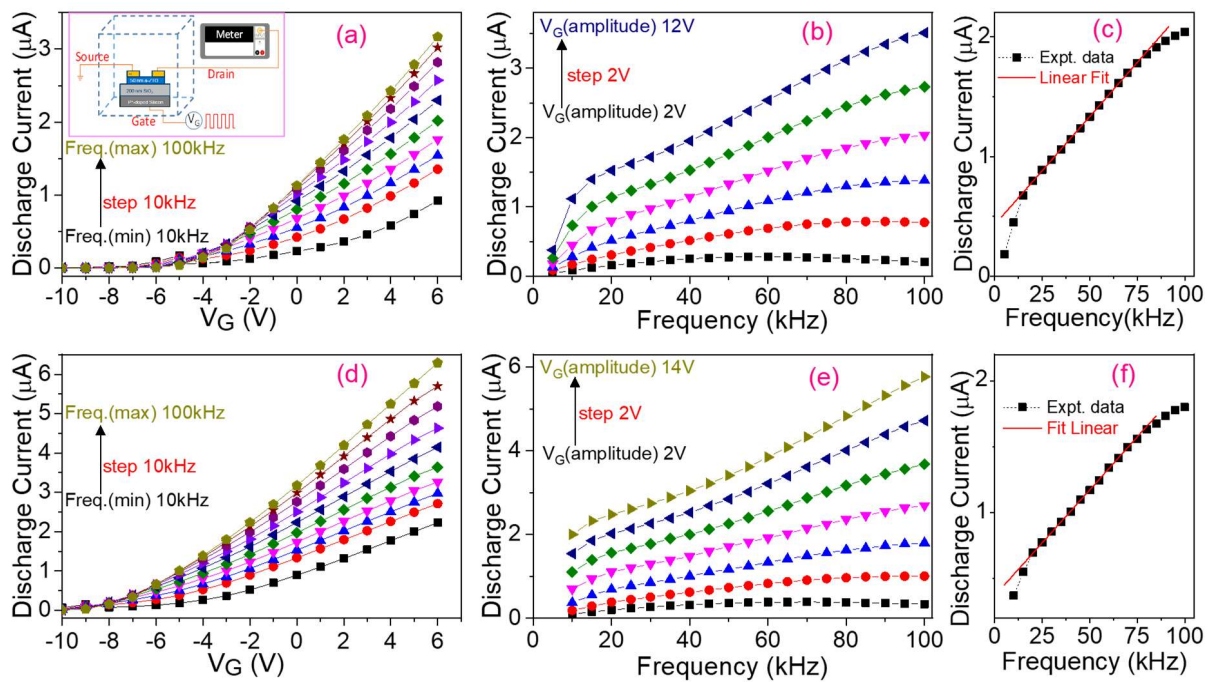

**Figure 6.** Measured discharge current data, as a function of gate voltage pulse for different pulse frequency (a) and gate pulse frequency for different voltage amplitude (b) for the sample 33ZTO TFT. The inset shows a schematic diagram of the overview of the discharge current analysis method with electrical connections. In (c), the representative linear fitting to the discharge current-gate pulse frequency experimental data. In (d-f) measured discharge current data of the same for another TFT, 50ZTO.
……………………………………………………………………………………………………………………...

In order to investigate further on the underlying physics of the observed photoresponse, we exploited the DCA technique[44,45] which is a modified charge pumping method (CPM)[46,47]. DCA is a powerful and effective method to extract the distribution of the DOS near the CB in oxide semiconductors in fabricated TFTs. A schematic diagram of the DCA measurement setup with circuit connection is shown as an inset in Fig. 6(a). In the DCA method, a periodic voltage pulse with 50% duty cycle at a given frequency is fed into the gate and the average discharge current is measured at the drain following a delay of one second. When the gate pulse amplitude is ramped up above $V_{FB}$, along with the population of the intrinsic bulk carriers (here electrons), the defect sites at the interface and in the active ZTO layer charge up. The charging times of the defect sites were found to be between µs and ms depending on the types of defect sites involved. When the gate pulse is ramped down, populated intrinsic carriers quickly discharge through drain and source. However, various trap sites in the sub-gap region of the active ZTO layer slowly discharge and release electrons over a time period of ms up to few seconds. Therefore, the tail portion of the discharge current is mainly associated with charge released from the sub-gap defect sites in the ZTO layer together with contributions from other trap sites in the ZTO-dielectric interface. This tail portion of the discharge current strongly depends on the applied pulse frequency, which shows a linear dependence in the low-frequency regime and then becomes saturated at a very high frequency. Using this method, a signature of the trap sites in the ZTO layer can be extracted from the slope of the discharge current-frequency data.

Figure 6 shows the measured discharge current from the 33ZTO and 50ZTO TFTs as a function of gate voltage and pulse frequency according to the DCA method, as discussed earlier. One can see that discharge current gradually increases as a function of gate voltage (Fig. 6(a and d)), and also as a function of frequency (Fig. 6(b and e)), as expected. Furthermore, a notable discharge current is observed only when there is a positive voltage with respect to the voltage $V_{FB}$ ($V_G$(amplitude) = $V_G$ - $V_{FB}$) applied to the gate. As the discharge current saturates at a pulse frequency of 120 kHz, the discharge current vs. frequency measurement was limited to 100 kHz. The trap/defects sites density is extracted from the slope, ($\partial I/\partial f$), in the linear region using the following equation,[44,45]

$$N_{defect\ site}(\#/cm^3) = \frac{2}{k} \cdot \frac{\partial I/\partial f}{V \cdot q} \qquad (3)$$

where $N_{defect\ site}$ represents the density of defect sites in the sub-gap region in the ZTO semiconductor, $k$ is the charge loss factor, $V$ is the active volume of the semiconducting layer and $q$ is the elemental charge. The slope is estimated from the linear fitting to the discharge current vs. frequency data for different $V_G$(amplitude) at a step of 1V, and a representative linear fitting for the 33ZTO and 50ZTO TFTs are shown in Fig. 6(c and f).

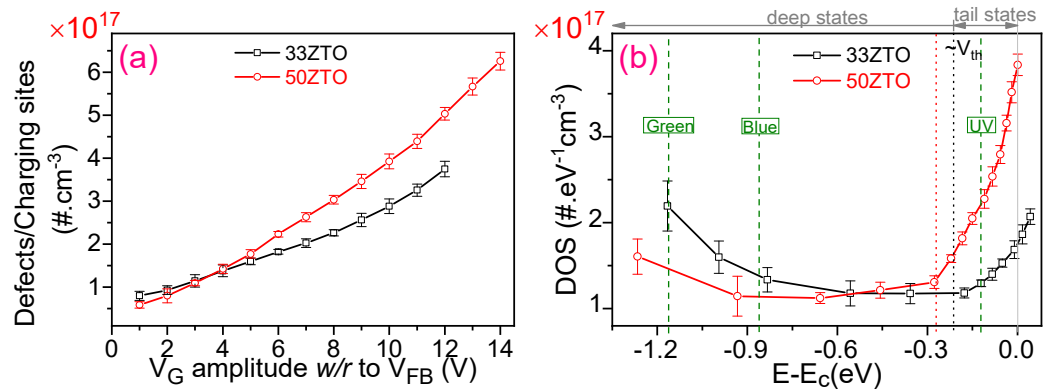

Figure 7. (a) The extracted sub-gap defects/charging sites density as a function of gate pulse voltage amplitude with reference to $V_{FB}$ for 33ZTO and 50ZTO TFTs. (b) The extracted density of sub-gap states (DOS) versus energy with reference to the conduction band minima. The green dashed lines represent photoexcitation energy levels for the 33ZTO at UV, blue and green light excitations.

…………………………………………………………………………………………………...

The extracted density of the sub-gap defect/charging sites for the oxide TFTs, as shown in Fig. 7(a) gradually increases when increasing the gate pulse amplitude. A defect density on the order of ~$10^{17}$/cm³ is obtained from both the TFTs, which is a relatively high density of defects compared to conventional semiconductors like silicon. The energy distribution of the sub-gap defects (DOS) in the sub-gap region is calculated by converting the gate amplitude to the surface potential energy. The surface potential, $\varphi_s$ as a function of $V_G$ is calculated by integrating the following equation,[48]

$$\varphi_s = \int_{V_{FB}}^{V_{GS}} \left(1 - \frac{C_g(V'_{GS})}{C_{ox}}\right) dV'_{GS} \qquad (4)$$

Figure 7(b) shows the extracted density profile of the sub-gap defect sites, which is exponentially distributed in energy, as expected. Interestingly, a significantly high density of tail states extending up to ~0.3 eV below the CB is observed in 50ZTO TFT. However, the oxygen vacancy related deep states defect density is higher in 33ZTO TFT than the 50ZTO TFT which is consistent with the XPS analysis results. We have previously reported the direct measurement of the density of sub-gap states by photothermal deflection spectroscopy (PDS), which is a very sensitive optical absorption method[49]. This measurement shows that the 33ZTO has a higher density of deep states than the 50ZTO. In contrast, 50ZTO film has a higher density of tail states (higher Urbach energy) than the 33ZTO. Even though the Urbach tail states by PDS method cannot distinguish between the conduction band tail and valence band tail, the density of sub-gap states measured by PDS is in agreement with the density of sub-gap states extracted by the DCA technique. As the transient photocurrent was measured in accumulation mode of the TFT at a gate voltage $V_{GS}<V_{FB}$, the photo-excited charge carriers are the sole source of observed high drain-to-source current. The photoexcitation energy levels of the UV, blue and green light excitations for the 33ZTO TFT ($E_g$=3.52 eV) are marked by green dashed lines in Fig. 7(b) at -0.12, -0.85 and -1.16 eV, respectively. In the case of 50ZTO TFT ($E_g$=3.66 eV), the reference levels are at -0.26, -0.99 and -1.30 eV for the UV, blue and green light excitations, respectively. If we compare the DOS and the generated photocurrent for both sets of TFTs, a higher DOS results in higher photogenerated current at a specific photoexcitation level. In the present case, under excitation by blue light, a 10.0 % increment of the DOS (deep states) in 33ZTO TFTs leads to a factor of four enhancement (~4.34) in the photocurrent compared to the 50ZTO TFT. Similarly, under excitation by UV light, an increment of 7.4 % of the DOS (tail states) in 50ZTO TFT leads to a factor of two enhancement (~2.63) in the photocurrent compared to the 33ZTO TFT. The sub-gap DOS extracted from the DCA technique thus provides direct evidence of the involvement of tail states in the ZTO when excited with UV light, and deep states when excited with blue/green light. Usually, the tail states are situated in the vicinity of the CB within 0.1 eV range. Involvement of the tail states is thus associated with high photosensitivity and strong PPC with long-lasting recovery time, while the deep states are associated with comparatively lower photosensitivity with mild PPC.

It is known that, upon photoexcitation with sufficient photon energy in oxide semiconductors, that the neutral oxygen vacancy ($V_o$) states are ionised and temporarily relocated near the CB[11,35,50]. The extracted sub-gap DOS profile reflects such singly ionised oxygen vacancy ($V_o^+$) states in the deep-acceptor states and doubly ionised oxygen vacancy ($V_o^{2+}$) states in tail-acceptor states. If such ionised oxygen vacancies solely contribute to the photoresponse process, one could expect higher photocurrent in the 33ZTO TFT than the 50ZTO TFT irrespective of the excitation photon energy, which is not seen here. This result indicates that in addition to the $V_o^{2+}$ states, there are some other activated defects present in the tail states in the 50ZTO, which are more optically irreversible than $V_o^{2+}$ states. It is widely considered that the oxygen vacancy defects are the major intrinsic origin of PPC effect in the oxide system. However, the effect of zinc interstitials or peroxide defects on the PPC is also proposed[51,52]. In the present case, in addition to the effect of oxygen vacancy on the PPC effect possibility of contribution from above mentioned two defects are verified. It is known that peroxide defects form under light illumination, only when a hydrogen-zinc vacancy defect complex (2H-$V_{Zn}$) present in the system. In the present case absence of H or OH attached to the metal or oxygen ions in the XPS data (spectra for O1s and Zn2p core levels) ruled out the possibility of involvement of peroxided defect in the observed long-lasting PPC. On the other hand, due to the high tin concentration, the probability of formation of additional zinc interstitial defects due to the substitutional doping error or lattice distortion is very high in 50ZTO. The XPS data also indicate presence of zinc interstitials in the 50ZTO. A first-principle theoretical calculation by Janotti *et al.*[43] on the ZnO system found a negative formation energy of the ionized $Zn_i^{2+}$ defects. As a result, neutral $Zn_i$ is easily ionized to $Zn_i^{2+}$ by releasing electrons and located near the CB edge[51]. This provides evidence that 50ZTO contains more tail states than the 33ZTO, and hence a high value of photosensitivity and long-lasting PPC. The ionisation-deionisation process of above two defects under light On and OFF conditions is shows schematically in the Fig. 8. Under illumination with photon energy close to the CB, the interstitial defects release electrons and contribute to the generated photocurrent along with $V_o$ defects. Just after switching off the light, the system remains in a positive bias condition at the same applied $V_{GS}$. Because, prolonged light illumination on the oxide TFTs shifts their characteristic $V_{th}$ towards the negative bias direction, known as light stress effect. In the present case, this shift is more than -16 V after 2 mins of light illumination. The $Zn_i$ subsequently return to their previous charge state by capturing free electrons. Like the $V_o$ defect, the $Zn_i$ is also a negative-U defect and its ionisation-deionisation process is more irreversible than the $V_o$ state which delays the recovery. Due to the high formation energy of the neutral $Zn_i$ defects, the deionisation process ($Zn_i^{2+}$ + $e^-$/$2e^- \rightarrow Zn_i^+/Zn_i$) during photocurrent decay is exceptionally slow. It is more likely, the $Zn_i^{2+}$ state captures a single electron and returns to $Zn_i^+$ state instead of directly returning to neutral $Zn_i$ state. This explains why an exceptionally slow decay is observed from the 50ZTO TFT under illumination with UV light. The photoresponse data (Fig. 4) and XPS data (Fig. 5) together further suggest that $Zn_i$ defects dominants over $V_o$ on the reset time when illuminated with light energy close to CB of the ZTO (tail states region). Therefore, combining the effects of $V_o$ and $Zn_i$ states provides a consistent explanation for the observed strong and long-lasting PPC in the 50ZTO TFT, where $V_o$ and $Zn_i$ states both are responsible for the long-lasting PPC.

## Conclusions

We have demonstrated that sub-bandgap light excitation near the CB of the ZTO photo-TFT leads to strong PPC with long-lasting recovery time (up to a month) with an associated semiconductor-to-conductor like transition. Furthermore, tin concentration in the ZTO ternary oxide modulates the photosensitivity as well as PPC recovery time. Under sub-bandgap excitation with UV light, photosensitivity reaches as high as ~$10^7$ and remains close to this value for a long time after switching off the light. Photon energy dependent photoresponse studies reveal that photoexcitation energies closer to the CB result in much higher recovery time. The sub-gap DOS has been experimentally extracted using the DCA technique. This result provides direct evidence of the involvement of sub-gap tail-states for strong PPC in ZTO, while deep-states contribute to mild PPC. The origin of the PPC in this material is attributed to ionised oxygen vacancy and ionised zinc interstitials, which require higher excitation energy. We conclude that, oxygen vacancy and zinc interstitials both are responsible for the observed long-lasting PPC effect. This conclusion is supported by the XPS analysis result. Such a high sensitivity with long-lasting photocurrent retention time in ZTO may find application in optical memory devices for holographic storage and non-volatile memory devices.

## Methods

**Device Fabrication.** We used a remote plasma reactive sputtering technique to prepare the ZTO thinfilms with 33 at.% and 50 at.% of tin with respect to zinc from zinc/tin alloy targets with different tin concentrations. Details of the preparation of the ZTO thinfilms and characterization results can be found elsewhere[23]. Bottom-gate, inverted staggered TFTs were fabricated by depositing an active ZTO channel layer of thickness ~50 nm on $SiO_2$ (film) on Si (100) substrate followed by air annealing at 500 °C. The $SiO_2$ layer was used as a gate dielectric and the heavily doped Si (100) substrate as the gate electrode. The top-contact was a thermally evaporated Al layer which served as the source/drain electrode. The active ZTO layer and the source/drain was patterned using a standard process for the fabrication of the TFTs with W= 1900 μm and L= 50 μm (W/L=38). In the final stage, the device chip was mounted on a stripboard and electrical connection to the external pin connectors was made by using gold wires (~25 μm thick) and a wire-bonder. Fabricated TFTs are marked as 33ZTO and 50ZTO for their active channel tin concentration of 33 at.% and 50 at.%, respectively.

**Characterizations and Electrical Measurements.** Elemental compositions and charging states of ions in the ZTO thinfilms were estimated from X-ray photoelectron spectroscopy (XPS) measurement (Multilab-2000, Thermo Scientific) using an Al $K_\alpha$ X-ray beam at 1486.7 eV in an ultra-high vacuum chamber. A standard UV-Vis spectrometer was employed to estimate bandgap energy of the ZTO thinfilms from the transmittance data.

All the electrical measurements of the TFTs were performed inside a light blocked dark box at room temperature using a computer controlled Keithley meter (6517B) and power supply (2230) through triax cables. The gate-to-source voltage was swept from -25 V to +25 V at a sweep rate of 0.33 V/s. The photoexcited electrical measurement was performed by illuminating the active area of the device with different narrow-band, high-power LEDs: UV (at 365 nm), blue (at 465 nm), green (at 526 nm) and red (at 662 nm) in ON and OFF conditions. Applied $V_{DS}$ and $V_{GS}$ voltages were 1 V and -7 V for the 33ZTO and 1 V and -11 V for the 50ZTO, respectively. For DCA measurement, a voltage pulse with frequency up to 100 kHz from a pulse generator (Tektronix AFG 2021) was applied to the gate and discharge current was measured at drain using Keithley meter while source connected to the ground. The pulse width and rising/falling time were fixed to 50% of the duty cycle at a given frequency and 100 ns, respectively.

## References

1. Park, S. H. K. *et al.* Transparent and photo-stable ZnO thin-film transistors to drive an active matrix organic-light-emitting diode display panel. *Adv. Mater.* **21**, 678-682 (2009).
2. Kamiya, T., Nomura, K. & Hosono, H. Present status of amorphous In−Ga−Zn−O thin-film transistors. *Sci. Technol. Adv. Mater.* **11**, 044305 (2010).
3. Jeon, S. *et al.* Nanometer-scale oxide thin film transistor with potential for high-density image sensor appliations. *ACS Appl. Mater. Interfaces* **3**, 1-6 (2011).
4. Yan, L. *et al.* The Development of High Mobility Zinc Oxynitride TFT for AMOLED. *SID Symp. Dig. Tech. Pap.* **46**, 769-771 (2015).
5. Lee, S. & Nathan, A. Subthreshold Schottky-barrier thin-film transistors with ultralow power and high intrinsic gain. *Science* **354**, 302-304 (2016).
6. Lim, T., Kim, H., Meyyappan, M. & Ju, S. Photostable $Zn_2SnO_4$ nanowire transistors for transparent displays. *ACS Nano* **6**, 4912-4920 (2012).
7. Nomura, K. *et al.* Room-temperature fabrication of transparent flexible thinfilm transistors using amorphous oxide semiconductors. *Nature* **432**, 488-492 (2004).
8. Sharp begins production of world's first LCD panels incorporating IGZO oxide semiconductors. (SHARP Corporation, http://www.sharp-world.com/corporate/news/120413.html, 2012).
9. Yoon, S. M. *et al.* Nonvolatile memory thin-film transistors using an organic ferroelectric gate insulator and an oxide semiconducting channel. *Semicond. Sci. Technol.* **26**, 034007 (2011).
10. Fortunato, E., Barquinha, P. & Martins, R. Oxide semiconductor thin-film transistors: a review of recent advances. *Adv. Mater.* **24**, 2945–2986 (2012).

11. Nathan, A., Lee, S., Jeon, S. & Robertson, J. Amorphous oxide semiconductor TFTs for displays and imaging. *J. Disp. Technol.* **10**, 917-927 (2014).
12. Young, D. L., Moutinho, H., Yan, Y. & Coutts, T. J. Growth and characterization of radio frequency magnetron sputter-deposited zinc stannate, $Zn_2SnO_4$, thin films *J. Appl. Phys.* **92**, 310-319 (2002).
13. Kim, C. H., Rim, Y. S. & Kim, H. J. The effect of a zinc–tin-oxide layer used as an etch-stopper layer on the bias stress stability of solution-processed indium–gallium–zinc-oxide thin-film transistors. *J. Phys. D: Appl. Phys.* **47**, 385104 (2014).
14. Chiang, H. Q., Wager, J. F., Hoffman, R. L., Jeong, J. & Keszler, D. A. High mobility transparent thin-film transistors with amorphous zinc tin oxide channel layer. *Appl. Phys. Lett.* **86**, 013503 (2005).
15. Chandra, R. D. *et al.* Tuning electrical properties in amorphous zinc tin oxide thin films for solution processed electronics. *ACS Appl. Mater. Interfaces* **6**, 773-777 (2014).
16. Sanctis, S. *et al.* Toward an understanding of thin-film transistor performance in solution-processed amorphous zinc tin oxide (ZTO) thin films. *ACS Appl. Mater. Interfaces* **9**, 21328–21337 (2017).
17. Cheng, B. *et al.* Individual Ohmic contacted $ZnO/Zn_2SnO_4$ radial heterostructured nanowires as photodetectors with a broad-spectral-response: injection of electrons into/from interface states. *J. Mater. Chem. C* **2**, 1808–1814 (2014).
18. Jiang, Q. J. *et al.* Ultraviolet photoconductivity of amorphous ZnAlSnO thin-film transistors. *RSC Adv.* **5**, 56116–56120 (2015).
19. Zhao, Y. *et al.* Band gap tunable $Zn_2SnO_4$ nanocubes through thermal effect and their outstanding ultraviolet light photoresponse. *Sci. Rep.* **4**, 6847 (2014).
20. Ghaffarzadeh, K. *et al.* Persistent photoconductivity in Hf-In-Zn-O thin film transistors. *Appl. Phys. Lett.* **97**, 143510 (2010).
21. Jeon, S. *et al.* Gated three-terminal device architecture to eliminate persistent photoconductivity in oxide semiconductor photosensor arrays. *Nature Mater.* **11**, 301-305 (2012).
22. Yang, B. S. *et al.* Improvement of the photo-bias stability of the Zn–Sn–O field effect transistors by an ozone treatment. *J. Mater. Chem.* **22**, 10994 (2012).
23. Niang, K. M., Cho, J., Heffernan, S., Milne, W. I. & Flewitt, A. J. Optimisation of amorphous zinc tin oxide thin film transistors by remote-plasma reactive sputtering. *J. Appl. Phys.* **120**, 085312 (2016).
24. Kanicki, J. & Martin, S. in *Thin-Film Transistors* (eds C. R. Kagan & P. Andry) 71-137 (Marcel Dekker Inc., 2003).
25. Maeng, J. *et al.* Effect of gate bias sweep rate on the electronic properties of ZnO nanowire field-effect transistors under different environments. *Appl. Phys. Lett.* **92**, 233120 (2008).
26. Du, Q. *et al.* Photo-assisted hysteresis of electronic transport for ZnO nanowire transistors. *Nanotechnol.* **29**, 115204 (2018).
27. Lee, H.-N., Song, B.-J. & Park, J. C. Fabrication of p-channel amorphous tin oxide thin-film transistors using a thermal evaporation process. *J. Disp. Technol.* **10**, 288-292 (2010).
28. Yao, J. *et al.* Electrical and photosensitive characteristics of a-IGZO TFTs related to oxygen vacancy. *IEEE Trans. Electron Devices* **58**, 1121-1126 (2011).
29. Cho, S. W., Yun, M. G., Ahn, C. H., Kim, S. H. & Cho, H. K. Bi-layer channel structure-based oxide thin-film transistors consisting of ZnO and Al-doped ZnO with different Al compositions and stacking sequences. *Electron. Mater. Lett.* **11**, 198-205 (2015).
30. Coleman, V. A. & Jagadish, C. in *Zinc oxide bulk, thin films and nanostructures: processing, properties, and applications* (eds Chennupati Jagadish & Stephen J. Pearton) (Elsevier, 2010).
31. Sanon, G., Rup, R. & Mansingh, A. Band-gap narrowing and band structure in degenerate tin oxide ($SnO_2$) films *Phys. Rev. B* **44**, 5672-5480 (1991).
32. Görrn, P., Lehnhardt, M., Riedl, T. & Kowalsky, W. The influence of visible light on transparent zinc tin oxide thin film transistors. *Appl. Phys. Lett.* **91**, 193504 (2007).
33. Ghaffarzadeh, K. *et al.* Instability in threshold voltage and subthreshold behavior in Hf–In–Zn–O thin film transistors induced by bias-and light-stress. *Appl. Phys. Lett.* **97**, 113504 (2010).
34. Dai, M. K., Liou, Y. R., Lian, J. T., Lin, T. Y. & Chen, Y. F. Multifunctionality of strong and long-lasting persistent photoconductivity: semiconductor–conductor transition in graphene nanosheets and amorphous InGaZnO hybrids. *ACS Photonics* **2**, 1057-1064 (2015).
35. Takechi, K., Nakata, M., Eguchi, T., Yamaguchi, H. & Kaneko, S. Comparison of ultraviolet photo-field effects between hydrogenated amorphous silicon and amorphous $InGaZnO_4$ thin-film transistors. *J. J. Appl. Phys.* **48**, 010203 (2009).
36. Lany, S. & Zunger, A. Anion vacancies as a source of persistent photoconductivity in II-VI and chalcopyrite semiconductors. *Phys. Rev. B* **72**, 035215 (2005).
37. Liu, C. *et al.* Tunable UV response and high performance of zinc stannate nanoparticle film photodetectors. *J. Mater. Chem. C* **4**, 6176--6184 (2016).
38. Chowdhury, M. D. H., Migliorato, P. & Jang, J. Temperature dependence of negative bias under illumination stress and recovery in amorphous indium gallium zinc oxide thin film transistors *Appl. Phys. Lett.* **102**, 143506 (2013).

39. Fernandes, C. et al. A sustainable approach to flexible electronics with zinc-tin-oxide thin-film transistors. *Adv. Electron. Mater.*, 1800032 (2018).
40. Kwoka, M. et al. XPS study of the surface chemistry of L-CVD $SnO_2$ thin films after oxidation. *Thin Solid Films* **490**, 36 – 42 (2005).
41. Chen, M. et al. X-ray photoelectron spectroscopy and auger electron spectroscopy studies of Al-doped ZnO films. *Appl. Surf. Sci.* **158**, 134–140 (2000).
42. Kim, H.-S. et al. Anion control as a strategy to achieve high-mobility and highstability oxide thin-film transistors. *Sci. Rep.* **3**, 1459 (2013).
43. Janotti, A. & Walle, C. G. V. d. Native point defects in ZnO. *Phys. Rev. B* **76**, 165202 (2007).
44. Jung, U. et al. Quantitatively estimating defects in graphene devices using discharge current analysis method. *Sci. Rep.* **4**, 4886 (2014).
45. Jung, U. et al. Quantitative analysis of interfacial reaction at a graphene/SiO2 interface using the discharging current analysis method. *Appl. Phys. Lett.* **104**, 151604 (2014).
46. Groeseneken, G., Maes, H. E., Beltran, N. & F., D. K. R. Reliable approach to charge-pumping measurements in MOS transistors. *IEEE Trans. Electron Devices* **ED-31**, 42–53 (1984).
47. Paulsen, R. E. & White, M. H. Theory and application of charge pumping for the characterization of Si-SiO$_2$ interface and nearinterface oxide traps. *IEEE Trans. Electron Devices* **41**, 1213–1216 (1994).
48. Wu, W.-J. et al. Analytical extraction method for density of states in metal oxide thin-film transistors by using low-frequency capacitance–voltage characteristics. *J. Disp. Technol.* **12**, 888-891 (2016).
49. Niang, K. M. et al. Zinc tin oxide thin film transistors produced by a high rate reactive sputtering: Effect of tin composition and annealing temperatures. *Phys. Status Solidi A* **214**, 1600470 (2017).
50. Liu, L. et al. p-type conductivity in N-doped ZnO: the role of the $N_{Zn}$– $V_O$ complex. *Phys. Rev. Lett.* **108**, 215501 (2012).
51. Khokhra, R., Bharti, B., Lee, H.-N. & Kumar, R. Visible and UV photo-detection in ZnO nanostructured thin flms via simple tuning of solution method *Sci. Rep.* **7**, 15032 (2017).
52. Kang, Y., Nahm, H.-H. & Han, S. Light-induced peroxide formation in ZnO: origin of persistent photoconductivity. *Sci. Rep.* **6**, 35148 (2016).

## Acknowledgments
This work was supported by the Engineering and Physical Sciences Research Council (EPSRC), United Kingdom, through research grants EP/M013006/1 and EP/M013650/1.

## Author Contributions
SD conceived the idea and performed all the electrical measurements and data analysis. KMN carried out thinfilm depositions and device fabrications. SD wrote the manuscript. SAL, AJF, and AN designed the research programme and were the principal investigators on the project. All the authors discussed results and reviewed the manuscript.

## Additional Information
**Competing Interests:** The author(s) declare no competing interests.